\documentstyle[11pt,newpasp,twoside,epsf]{article}
\def\edcomment#1{\iffalse\marginpar{\raggedright\sl#1\/}\else\relax\fi}
\reversemarginpar

\begin{document}
\title{The near-IR luminosity-metallicity relationship 
for dwarf irregular galaxies}
\author{Ivo Saviane}

\affil{European Southern Observatory, 3107 A. de Cordova, Santiago, Chile}

\author{R. Riegerbauer$^{1,2}$, E. V. Held$^3$, V. Ivanov$^1$, D.
Alloin$^1$, F. Bresolin$^4$, Y. Momany$^5$, R. M. Rich$^6$,
L. Rizzi$^{3,4,5}$}
\affil{$^1$ ESO, $^2$Lule{a} University of Technology, $^3$OAPD, 
$^4$IfA Hawaii, $^5$UniPD, $^6$UCLA}

\begin{abstract}
We briefly describe our on-going investigation of the near-IR
luminosity-metallicity relationship for dwarf irregular galaxies in
nearby groups of galaxies. The motivations of the project and the
observational databases are introduced, and a preliminary result is
presented. The $12+\log \rm (O/H)$ vs. $H$ plane must be populated with
more low-luminosity galaxies before a definite conclusion can be drawn.
\end{abstract}

\section{The Project}

Although early studies 
found a very well defined luminosity-metallicity (L-Z) relationship for
dwarf irregular (dIrr) galaxies (Skillman, Kennicutt, \& Hodge 1989;
Richer \& McCall 1995), more recent investigations seem to suggest
a mild
relationship with much scatter, or no relationship at all (Hidalgo-Gamez
\& Olofsson 1998; 
Hunter \& Hoffman 1999; Skillman, C\^{o}t\'e, \& Miller 2003, hereafter
SCM03).
The main difficulties with the present status 
are that 
(a) all the studies use blue luminosities as tracers of the mass,
(b) galaxies in different environments are used,
(c) inhomogeneous data are used, and 
(d) the scatter in distance modulus is unknown.
In order to overcome these limitations, we started a medium-term project
whose aims are 
(i) collecting homogeneous samples of oxygen abundances,
(ii) focusing on well-defined environments with reduced
distance range,
and (iii) collecting near-IR luminosities for all the objects.
As a first step, we are collecting data for the three 
nearest groups of galaxies:
the M\,81, Sculptor, and Cen~A groups at a distance of about 
$3.5$, $2.5$, and $3.5$~Mpc, respectively.  
The interaction rate 
in the M\,81 and Cen~A groups is higher
than that in Sculptor, so our targets offer the chance to test the effect
of the environment on the L-Z relationship.

\section{First Results}

Spectroscopy  has been obtained 
at the Lick observatory
with KAST@3.5m for the M\,81 group, 
and at ESO with EFOSC2@3.6m for the Sculptor group. 
Near-IR imaging has been obtained
at La Palma observatory 
with  INGRID@WHT, and  at ESO with SOFI@NTT for the M\,81 and Sculptor
groups, respectively.
Observations of the Sculptor group are presented here.
The reductions of the M\,81 group are in progress, and observations of the
Cen~A group 
have been planned.

\begin{figure}[t]

\centerline{{\centering \leavevmode
\epsfxsize=0.70\textwidth \epsfbox{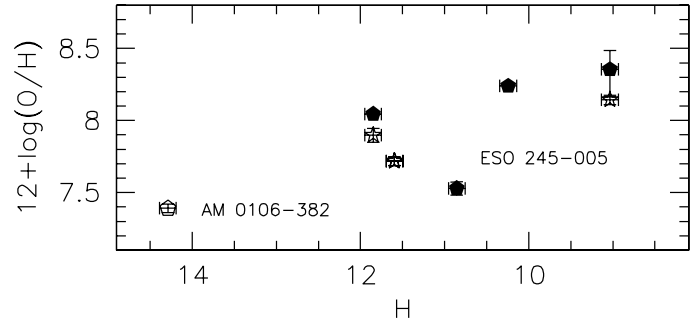}}}

\caption{Oxygen abundance vs. $H$ magnitude for dIrrs of the
Sculptor group in our sample (open and filled pentagons) and of that of
SCM03 (stars). The open pentagon identifies an oxygen abundance computed
via the empirical method. }
\end{figure}

Oxygen abundances 
are plotted vs. $H$ magnitudes in Fig.~1.  Note that 
the observed H{\sc ii} region of ESO
245-005 lies in the outskirts of the galaxy, 
so the
abundance of the central 
star-forming regions might be closer to the trend defined
by the other galaxies. The trend is rather well defined, although with
some scatter.  
On the other hand, the existence of the relationship depends very much
on the single point defined by AM 106-382, whose abundance has been
found via the empirical method (Edmunds \& Pagel 1984). More data 
are clearly needed, in particular for low-luminosity galaxies.
The abundances of the objects from SCM03
agree with our determinations, reinforcing 
the correlation.
The figure shows the advantage of working in the near-IR: the galaxies
cover almost $6$ magnitudes in $H$, while the SCM03 sample spans only $3$
magnitudes in $B$. The impact of the unknown individual distance moduli is
then reduced in the near-IR.

\end{document}